\documentstyle[12pt]{article}

\newcommand{\bbc}{\begin{center}}
\newcommand{\eec}{\end{center}}

\begin{document}
\begin{flushright}
Alberta Thy-16-99 \\
Sept. 1999\\
\end{flushright}

\vspace {0.3in}

\begin{center}
\Large\bf Resonant final-state interactions in $D^0 \rightarrow \bar{K}^{0} {\eta}, \bar{K}^{0} {\eta^{\prime}}$ Decays\\
\vspace {0.3in}
{\large El hassan  El aaoud and A. N. Kamal } \\ \vspace {0.1in}
 \small\em Theoretical Physics Institute and Department of Physics, \\
 \small\em University of Alberta, Edmonton, Alberta T6G 2J1, Canada.
\end{center}
\vspace {0.1in}

\begin{abstract}
We have investigated  the effect of the isospin ${1 \over 2}$, $J^P = 0^+$ resonant state $K^*_0(1950)$ on the decays $D^0 \rightarrow \bar{K}^{0} \eta$ and $D^0 \rightarrow \bar{K}^{0} \eta^\prime$ as a function of the branching ratio sum $ r =Br(K^*_0(1950)\rightarrow \bar{K}^{0} {\eta}) + Br(K^*_0(1950)\rightarrow \bar{K}^{0} {\eta^\prime})$ and coupling constants $g_{K^*_0\bar{K}^0\eta}$, $g_{K^*_0\bar{K}^0\eta^\prime}$. We have used a factorized input for $D^0 \rightarrow K^*_0(1950)$ weak transition through a $\pi K$ loop. We estimated both on- and off-shell contributions from the loop. Our calculation shows that the off-shell effects are significant. For $r\geq 30\%$ a fit to the decay amplitude $A(D^0 \rightarrow \bar{K}^0 \eta^\prime)$ was possible, but the amplitude $A(D^0 \rightarrow \bar{K}^0 \eta)$ remained at its factorized value. For small values of $r$, $r\leq18\%$, we were able to fit $A(D^0 \rightarrow \bar{K}^0 \eta)$, and  despite the fact that $A(D^0 \rightarrow \bar{K}^0 \eta^\prime)$ could be raised by almost $100\%$ over its factorized value, it still falls short of its experimental value. A simultaneous fit to both amplitudes $A(D^0 \rightarrow \bar{K}^0 \eta^\prime)$ and $A(D^0 \rightarrow \bar{K}^0 \eta)$ was not possible. We have also determined the strong phase of the resonant amplitudes for both decays.  
\end{abstract}

\bbc (PACS numbers:13.25.Ft,  13.25.-k, 14.40.Lb) \eec
\newpage

\bbc 
{\bf I. INTRODUCTION}
\eec

A common method of evaluating the matrix  elements of two-body hadronic decays of heavy mesons, $B$ and $D$, is based on the factorization approximation \cite{ref:c3bsw} which utilizes model form factors. However, this approximation has had only limited success in describing two-body hadronic decays of the $D$ meson \cite{ref:c3KK, ref:c3ksuv, ref:c3BLA, ref:c3VKK, ref:c3verma}. In particular, the factorization approximation not only underestimates the decay rates for $D^0 \rightarrow \bar{K}^0 \eta$ and $D^0 \rightarrow \bar{K}^0 \eta^\prime$, it generates $\Gamma(D^0 \rightarrow \bar{K}^0 \eta) > \Gamma(D^0 \rightarrow \bar{K}^0 \eta^\prime)$ in contradiction with the experiment \cite{ref:c3pdg98}. In an attempt to remedy this discrepancy, Ref. \cite{ref:c3VKK} studied the above decays in the factorization approximation but included the annihilation term. They found that unlikely large form factors for $ K \rightarrow \eta (\eta^\prime)$ transitions were required in order to bridge the gap between theory and experiment. Ref. \cite{ref:c3verma} on the other hand introduced nonfactorized contributions and used a flavor-$SU(3)$ parametrization for the nonfactorized matrix elements to fit the data. Their conclusions imply a large value for the hair-pin amplitude.

Hadronic decays of mesons are complicated by the presence of final-state strong interactions ($FSI$) between hadrons in the final state. The importance of $FSI$ in hadronic decays of $D$ meson has been known for a long time; its role was emphasized by several authors  \cite{ref:c3lipkam} almost twenty years ago. The long-range $FSI$ generates phases in the decay amplitudes \cite{ref:c3kam} and the most dramatic effect of $FSI$ is induced by the interference between different isospin amplitudes which depends on the phase difference \cite{ref:c3bsw, ref:c3lipkam, ref:c3wiss} between different isospin amplitudes. In decays with a single isospin final states, as in $ D^0 \rightarrow \bar{K}^0 \eta (\eta^\prime)$, isospin interference effects are absent. However, $FSI$ also leads to a change in the magnitude of the decay amplitude, and not simply a rotation in the complex amplitude plane. Hence we expect $FSI$ 
to effect the decay rates in single-isospin channel decays too.

\begin{figure}

%%Begin InstantTeX Picture
\let\picnaturalsize=N
\def\picsize{4.0in}
\def\picfilename{fig31.eps}
%If you do not have the picture file add:
%\let\nopictures=Y
%to the beginning of the file.
\ifx\nopictures Y\else{\ifx\epsfloaded Y\else\input epsf \fi
\let\epsfloaded=Y
\centerline{\ifx\picnaturalsize N\epsfxsize \picsize\fi \epsfbox{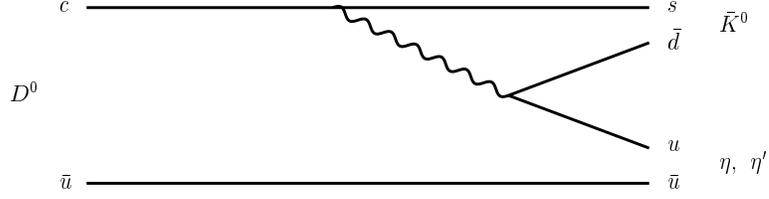}}}\fi
%%End InstantTeX Picture

\caption{Quark diagram contributing to the factorized amplitude $A^f$ for $D^0 \rightarrow \bar{K}^{0} {\eta}, \bar{K}^{0} {\eta^\prime}$.}
\label{fig:c3.1}
\end{figure}

\begin{figure}

%%Begin InstantTeX Picture
\let\picnaturalsize=N
\def\picsize{4.0in}
\def\picfilename{fig32.eps}
%If you do not have the picture file add:
%\let\nopictures=Y
%to the beginning of the file.
\ifx\nopictures Y\else{\ifx\epsfloaded Y\else\input epsf \fi
\let\epsfloaded=Y
\centerline{\ifx\picnaturalsize N\epsfxsize \picsize\fi \epsfbox{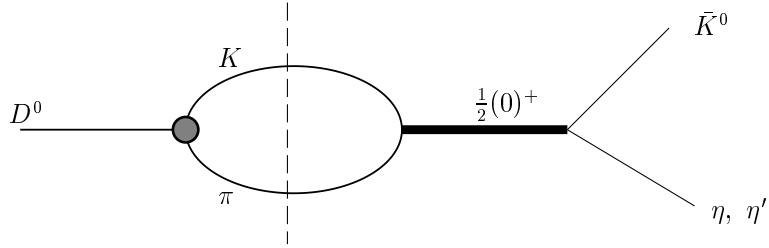}}}\fi
%%End InstantTeX Picture
\caption{Resonant contribution to $D^0 \rightarrow \bar{K}^{0} {\eta}, \bar{K}^{0} {\eta^\prime}$. The vertical dashed line represents the cut when the particles in the loop are on-shell. The thick line represents the resonance $K^*_0(1950)$, and the shaded circle represents the weak vertex ${\cal V}(k^2)$.}
\label{fig:c3.2}
\end{figure}

In this paper we have considered the $FSI$ effect of $K^*_0(1950)$ resonance on $D^0 \rightarrow \bar{K}^0 \eta$ and $D^0 \rightarrow \bar{K}^0 \eta^\prime$ decays. The mechanism we are proposing for resonant $FSI$ is as follows: $D^0 \rightarrow \bar{K}^0 \eta(\eta^\prime)$ are color-suppressed decays in the factorization approximation ( see Fig. \ref{fig:c3.1}). The resonance $K^*_0(1950)$ has a substantial branching ratio $(\sim 50\%)$ to $K \pi$ mode, leaving room for its coupling to $K \eta ( \eta^\prime)$ channels. We propose that the effect of $K^*_0(1950)$ on $D^0 \rightarrow \bar{K}^0 \eta (\eta^\prime)$ could be estimated via the Feynman diagram shown in Fig. \ref{fig:c3.2} where in the loop we include both $K^-\pi^+$ (color-favored decay) and $\bar{K}^0\pi^0$ (color-suppressed decay) states. Such mechanism has been invoked in Ref. \cite{ref:c3dai} in $D^0 \rightarrow \bar{K}^0 K^0$ decay. However, in contrast to Ref. \cite{ref:c3dai}, where only the on-shell loop contribution is retained. We evaluate both on- and off-shell loop contributions in Fig. \ref{fig:c3.2}. 

Recently, Gronau \cite{ref:c3gro} has also discussed the role of resonant FSI on $D$ decays. We relegate a discussion of these works \cite{ref:c3dai, ref:c3gro}, and their relationship to ours, to the last section of this paper.

In Sec. II we detail the model and the method of calculation. The results are presented in Sec. III, and Sec. IV deals with the discussion.

\vskip 5mm  
\begin{center}
{\bf II.  METHOD OF CALCULATION}
\end{center}
\begin{center}
{\bf A. Calculation without final-state interactions}
\end{center}

The decays $D^0 \rightarrow \bar{K}^{0} {\eta}, \bar{K}^{0} {\eta^\prime}$  are Cabibbo-favored and are induced by the effective weak Hamiltonian which can be reduced to the following color-favored ($CF$) and color-suppressed ($CS$) forms \cite{ref:c3ksuv} $( {\tilde{G}} = \frac{G_{F}}{\sqrt{2}} V_{cs} V^{*}_{ud} )$:
\begin{eqnarray}
{H }_{CF} &=&  {\tilde{G}} [ a_{1} ( \bar{u} d)( \bar{s} c)  + {c}_{2} {O}_{8}],  \nonumber \\
{H}_{{CS}} &=&  {\tilde{G}} [ a_{2} ( \bar{u} c)( \bar{s} d)  + {c}_{1} \tilde{O}_{8}],
\label{eq:c3ham}
\end{eqnarray}
where $V_{qq'}$ are the CKM matrix elements. The brackets $(\bar{u}d)$ etc. represent ($V$ - $A$) color-singlet Dirac bilinears. ${O}_{8}$ and ${\tilde{O}}_{8}$ are products of color-octet currents:  $ O_8 = \frac{1}{2}\sum_{a=1}^{8}{(\bar{u}{\lambda}^{a}d)(\bar{s}{\lambda}^{a}c})$ and ${\tilde{O}}_{8} = \frac{1}{2}\sum_{a=1}^{8}{(\bar{u}{\lambda}^{a}c)(\bar{s}{\lambda}^{a}d})$ where ${\lambda}^{a}$ are the Gell-mann matrices. The parameters $a_1$ and $a_2$, related to the Wilson coefficients, are chosen to be $a_1 =1.26\pm 0.04$ and $a_2 = -0.51 \pm 0.05$ \cite{ref:c3ksuv}. Note that these values of $a_1$ and $a_2$ are achieved in $N \rightarrow \infty$ limit, where $N$ is the number of colors.

In the factorization approximation one neglects the contribution from ${O}_{8}$ and ${\tilde{O}}_{8}$, and the  matrix element of the first term is written as a product of  two current matrix elements. Since we are effectively working with $N \neq 3$, one could argue that the nonfactorization arising from ${O}_{8}$ and ${\tilde{O}}_{8}$ is being included since working with $N \neq 3$ is equivalent to working with $N = 3$ but including nonfactorization effects. The quark diagram which contributes to $D^0 \rightarrow \bar{K}^{0} {\eta}({\eta^\prime})$ decays is shown in Fig. \ref{fig:c3.1}. The physical particles $\eta$ and $\eta^\prime $ are mixtures of the flavor-singlet $\eta_0 = {1 \over \sqrt{3} }| u\bar{u} + d\bar{d} + s\bar{s}>$ and the flavor-octet
$\eta_8 = {1 \over \sqrt{6} }| u\bar{u} + d\bar{d} - 2s\bar{s}>$ with a mixing angle $\theta_p = - 20^{\circ}$ \cite{ref:c3pdg96}:
\begin{eqnarray}
\eta &=& \eta_8\cos{\theta_p} -\eta_0\sin{\theta_p} \\
\eta^\prime &=& \eta_8\sin{\theta_p} +\eta_0\cos{\theta_p}.
\label{eq:c3mix}
\end{eqnarray}
The factorized amplitude for the decay $D^0 \rightarrow \bar{K}^{0} {\eta} ({\eta^\prime})$ is given by (superscript $f$ refers to 'factorized')

\begin{equation}
A^f(D^0 \rightarrow \bar{K}^{0} \eta) =  {\tilde{G}}a_2 
\left\langle\bar{K}^0 \mid\bar{s} d\mid 0 
\right\rangle\left\langle \eta \mid \bar{u} c \mid D^0\right\rangle.
\label{eq:c3facamp}
\end{equation}
In calculating the decay amplitude in Eq. (\ref{eq:c3facamp}), we use the following definitions,
\begin{eqnarray}
\langle K(p) |  (\overline{s} d)_\mu | 0 \rangle & = & - i f_K p_\mu \\
\langle \eta (p_\eta) |  (\overline{u} c)_\mu | D(p_D) \rangle & = & \left( p_D + p_\eta -
\frac{m^2_D - m^2_\eta}{q^2} q \right)_\mu F^{D\eta}_1(q^2)  \nonumber \\
&&+ \frac{m^2_D - m^2_\eta}{q^2} q_\mu F^{D\eta}_0(q^2)  
\label{eq:c3decdef} 
\end{eqnarray}
where $ q = p_D -p_{\eta}$ is the momentum transfer, $f_K$ is the decay 
constant of the $K$ meson, $ F^{D\eta}_i(q^2), (i = 0,1)$ are invariant form factors 
defined in \cite{ref:c3bsw1}. The factorized amplitudes for $D^0 \rightarrow \bar{K}^{0} \eta (\eta^\prime)$ are  
\begin{equation}
\left [\matrix{
A^f(D^0 \rightarrow \bar{K}^{0} \eta)\cr
A^f(D^0 \rightarrow \bar{K}^{0} \eta^\prime)\cr
}\right ]  = -i  {\tilde{G}}{ a_2\over \sqrt{2}} f_K \left [\matrix{
\sin{\theta^\prime}(m_D^2 - m_\eta^2) F_0^{D\eta}(m_K^2)\cr
\cos{\theta^\prime} (m_D^2 - {m_{\eta}^\prime}^2) F_0^{D\eta^\prime}(m_K^2)\cr
}\right ],
\label{eq:c3decamp}
\end{equation}
\\
where $\theta^\prime$ is given by
\begin{eqnarray}
\sin{\theta^\prime} &=&{1 \over \sqrt{3} }\cos{\theta_P} - \sqrt{{2 \over 3}} \sin{\theta_P}\nonumber \\ 
\cos{\theta^\prime} &=&\sqrt{{2 \over 3}} \cos{\theta_P} + \sqrt{{1 \over 3}} \sin{\theta_P}.
\end{eqnarray}
The corresponding decay rates are given by
\begin{equation}
\Gamma^f(D^0 \rightarrow \bar{K}^{0} \eta (\eta^\prime)) = \mid A^f\mid^2{\mid {\bf p}\mid \over 8\pi m_D^2},
\label{eq:c3decf}
\end{equation}
where $\mid {\bf p}\mid$ is the center of mass momentum and $m_D$ is the mass of the decaying $D$ meson. \\

\begin{center}
{\bf B. Calculation with resonant final-state interactions}
\end{center}

The resonances contributing to $D^0 \rightarrow (\bar{K}^{0} {\eta}, \bar{K}^{0} {\eta^\prime})$ must have isospin and spin-parity assignment $I(J^P) ={1 \over 2} (0^+)$. There are two particles, $\bar{K}^*_0(1950)$ and $\bar{K}^*_0(1430)$, with such properties \cite{ref:c3pdg98}: $m_1 = m_{\bar{K}^*_0} = 1945 \pm 10 \pm 20 MeV$, $\Gamma_1 = 201 \pm 34 \pm 79 MeV$ and $m_2 = m_{\bar{K}^*_0} = 1429 \pm 4 \pm 5 MeV$, $\Gamma_2 = 287 \pm 10 \pm 21 MeV$. Although $m_2 + {\Gamma_2 \over 2}$ is much smaller than $m_D$ and $\bar{K}^*_0(1430)$ decays almost exclusively to $\pi K$ channel, we cannot prejudice its effect to be insignificant in $D^0 \rightarrow (\bar{K}^{0} {\eta}, \bar{K}^{0} {\eta^\prime})$ \cite{ref:c3EK}. However, in this paper we consider the contribution from $\bar{K}^*_0(1950)$ only.

The contribution of $\bar{K}^*_0(1950)$ to $D^0 \rightarrow \bar{K}^0 \eta (\eta^\prime)$ is  represented by Feynman diagram shown in Fig. \ref{fig:c3.2}, where in the loop we consider both the color-favored state $K^-\pi^+$ and the color-suppressed state $\bar{K}^0\pi^0$. Evaluation of the diagram in Fig. \ref{fig:c3.2} gives the following amplitudes  ( superscript $r$ and subscript $-+$, $00$ refer to 'resonant' and $K^-\pi^+$, $\bar{K}^0 \pi^0$ intermediate state, respectively)
\begin{eqnarray}
A^r_{-+}  &=&  {i \over (2\pi)^4}\int {d^4k{ {\cal V}^{(1/2)}_{-+}(k^2) \over (k^2 - m_\pi^2)((w-k)^2 - m_K^2)}} \times A_{-+}^{strong} \nonumber \\
&\equiv& I_{-+}\times A^{strong}_{-+},
\label{eq:c3a-+}
\end{eqnarray}
with
\begin{equation}
A_{-+}^{strong}  = g_{-+}{1 \over (m_D^2 - m_R^2) + i\Gamma m_R}f,
\label{eq:c3astro-+}
\end{equation}
and
\begin{eqnarray}
A^r_{00}  &=&  {i \over (2\pi)^4}\int {d^4k{ {\cal V}^{(1/2)}_{00}(k^2) \over (k^2 - m_\pi^2)((w-k)^2 - m_K^2)}} \times A_{00}^{strong} \nonumber \\
&\equiv& I_{00}\times A^{strong}_{00},
\label{eq:c3a00}
\end{eqnarray}
with
\begin{equation}
A_{00}^{strong}  = g_{00}{1 \over (m_D^2 - m_R^2) + i\Gamma m_R}f ,
\label{eq:c3astro00}
\end{equation}
where $g_{-+}$ and $g_{00}$ are the couplings of $K^-\pi^+$ and $\bar{K}^0\pi^0$ states to $\bar{K}^*_0(1950)$ and $f$ is the coupling of $\bar{K}^0\eta (\eta^\prime)$ state to $\bar{K}^*_0(1950)$, $m_R$ is the resonance mass, $\Gamma$ its width, $w$ is the four momentum of the decaying particle $(w = (m_D,0)$ in the C.M.) and $k$ is the loop momentum to be integrated over; ${\cal V}^{(1/2)}_{-+}(k^2)$ and ${\cal V}^{(1/2)}_{00}(k^2)$ are the vertex functions in isospin 1/2 state. They are related to the amplitudes $A(D^0 \rightarrow K^- \pi^+)$ and $A(D^0 \rightarrow  \bar{K}^0\pi^0)$ and are evaluated in the following.

Although $A^r_{-+}$ and $A^r_{00}$ get contributions from color-favored and color-suppressed intermediate states respectively, they are not independent but are related by isospin and $SU(3)$ symmetry as the following analysis elucidates. As the resonance $K^*_0(1950)$ has isospin $1/2$ and as strong interactions conserve isospin, only the isospin component $A^{(1/2)}$ in the following will contribute to ${\cal V}^{(1/2)}_{-+}(k^2)$ and ${\cal V}^{(1/2)}_{-+}(k^2)$. The isospin decomposition of the decay amplitudes,
\begin{eqnarray}
A(D^0 \rightarrow K^-\pi^+) &=&\sqrt{{1 \over 3}}A^{(3/2)} + \sqrt{{2 \over 3}} A^{(1/2)} \nonumber \\
A(D^0 \rightarrow \bar{K}^{0} \pi^0) &=& \sqrt{{2 \over 3}}A^{(3/2)} - \sqrt{{1 \over 3}} A^{(1/2)},
\label{eq:c3isodecom}
\end{eqnarray}
allows us to calculate $A^{(1/2)}$ which is needed as the input. For the color-favored  intermediate state $K^- \pi^+$ we get from Eq. (\ref{eq:c3isodecom})
\begin{eqnarray}
{\cal V }^{(1/2)}_{-+}(k^2) &=&\sqrt{{2 \over 3}} A^{(1/2)} \nonumber \\ 
 &=& \sqrt{{2 \over 3}} \left\{\sqrt{{2 \over 3}}A(D^0 \rightarrow K^- \pi^+ ) - \sqrt{{1 \over 3}} A(D^0 \rightarrow K^0\pi^0)\right\},
\label{eq:c3iso-+}
\end{eqnarray}
while for the color-suppressed intermediate state, we get
\begin{eqnarray}
{\cal V }^{(1/2)}_{00}(k^2) &=&-\sqrt{{ 1\over 3}} A^{(1/2)} \nonumber \\ 
 &=& -\sqrt{{1 \over 3}} \left\{\sqrt{{2 \over 3}}A(D^0 \rightarrow K^- \pi^+ ) - \sqrt{{1 \over 3}} A(D^0 \rightarrow K^0\pi^0)\right\}.
\label{eq:c3iso00}
\end{eqnarray}
Therefore, isospin symmetry offers the following relation between the two vertex functions ${\cal V}^{(1/2)}_{-+}(k^2)$ and ${\cal V}^{(1/2)}_{00}(k^2)$
\begin{equation}
{\cal V}^{(1/2)}_{-+}(k^2) = -\sqrt{2} {\cal V}^{(1/2)}_{00}(k^2).
\label{eq:c3vv}
\end{equation}
Let us now turn to the coupling of $K^*_0(1950)$ to mesons. Using a $SU(3)$-invariant strong Hamiltonian which couples a scalar octet $S$ to two pseudoscalars octets $P$,
\begin{equation}
H_{SPP} = {g \over 2}Tr(\{P^T,P^T\}S),
\label{eq:c3Hspp}
\end{equation}
we obtain 
\begin{equation}
g =g_{K_0^*K^-\pi^+}= g_{-+},~~~~ g_{K_0^*K^0\pi^0} = g_{00} = - {g \over \sqrt{2} }, ~~~~ g_{K_0^*\bar{K}^0\eta_8} = -{g \over \sqrt{6} }.
\label{eq:c3g8}
\end{equation}
From equations (\ref{eq:c3astro-+}), (\ref{eq:c3astro00}) and (\ref{eq:c3g8}) the $SU(3)$ symmetry gives the following relation between the strong amplitudes $A^{strong}_{-+}$ and $A^{strong}_{00}$:
\begin{equation}
A^{strong}_{-+} = -\sqrt{2} A^{strong}_{00}.
\label{eq:c3aas}
\end{equation}
Hence, using equations (\ref{eq:c3a-+}), (\ref{eq:c3a00}), (\ref{eq:c3vv}) and (\ref{eq:c3aas}) the resonant amplitudes $A^r_{-+}$ and $A^r_{00}$ are related by
\begin{equation}
A^r_{-+} = 2 A^r_{00}
\label{eq:c3aar}
\end{equation}
and the total resonant amplitude $A^r$ is
\begin{eqnarray}
A^r &=&A^r_{-+} + A^r_{00} \nonumber \\
&=&{3 \over 2}A^r_{-+}.
\label{eq:c3arsym}
\end{eqnarray}
To proceed further with the calculation models and approximations are used.

For an explicit calculation of the amplitude $A^r_{-+}$ in Eq. (\ref{eq:c3a-+}) we need to calculate the integral $I_{-+}$ for which we require the momentum dependence of the vertex function ${\cal V}^{(1/2)}_{-+}(k^2)$. For this we assume the form dictated by the factorization assumption but with both $\pi$ and $K$ not necessarily on their mass-shells (see Ref. \cite{ref:c3dgh}),  
\begin{eqnarray}
A^f(D^0 \rightarrow K^-(p) \pi^+(k)) &=& -i  {\tilde{G}}a_1 f_\pi (m_D^2 - p^2) F_0^{DK}(k^2) \nonumber \\
A^f(D^0 \rightarrow \bar{K}^{0}(p) \pi^0(k)) &=& -i  {\tilde{G}}{a_2 \over \sqrt{2} } f_K (m_D^2 - k^2) F_0^{D\pi}(p^2).
\label{eq:c3interdec}   
\end{eqnarray}
From Eqs. (\ref{eq:c3iso-+}) and (\ref{eq:c3interdec}) we then obtain 
\begin{eqnarray}
{\cal V}^{(1/2)}_{-+}(k^2) &=&-i \tilde{G} \sqrt{{2 \over 3}} \left\{\sqrt{{2 \over 3}}a_1 f_{\pi}(m_D^2 - (w-k)^2)F_0^{DK}(k^2) \right. \nonumber  \\ 
&&\left. - {a_2\over \sqrt{6}} f_K(m_D^2 - k^2)F_0^{D\pi}((w - k)^2)\right\},
\label{eq:c3vertex}  
\end{eqnarray}
where $p$ and $k$ are four-momenta of $K$ and $\pi$ mesons in the loop, respectively. They are related by momentum conservation at the vertex, $ w = p + k$. Since in our calculation, the intermediate particles $ (K,\pi)$ are allowed to be off-shell, we have to assume a behavior of the form factors  in the vertex function, Eq. (\ref{eq:c3vertex}), as the particles go off-shell. Form factors with a dipole dependence: $ F(k_X^2) = \left( {\lambda^2 - m_X^2 \over \lambda^2 - k_X^2}\right)^2$ (where $\lambda$ is fixed by experiment) have been used in the past \cite{ref:c3off} to describe off-shellness of intermediate state particles. In this work we have used the following phenomenological form factor
\begin{equation}
F_0^{DX}(k^2) ={ F_0^{DX}(0)\over (1 - {k^2 \over\Lambda^2})( 1 - {(w-k)^2 - m_X^2 \over \Lambda^2})}. 
\label{eq:c3ffdipo}
\end {equation}
In principle, one could use two different mass scales, $\Lambda_1$ and $\Lambda_2$. We chose to work with a single mass scale for simplicity\footnote{The mass scale $\Lambda$ in the factor $F_0^{DK}$ is different from the $\Lambda$ in the form factor $F_0^{D\pi}$.}. Beside having a dipole dependence, this form factor satisfies the following limit: As $K$ goes on-shell ( $p^2 = (w-k)^2 \rightarrow  (m_X^2 = m_K^2)$ the form factor in Eq. (\ref{eq:c3ffdipo}) reduces to the usual monopole form, 
\begin{equation}
 F_0^{DK}(k^2)= {F_0^{DK}(0) \over 1- {k^2 \over \Lambda^2}},
\label{eq:c3ffmono}
\end{equation}
with the pole mass  $\Lambda$ given in Ref. \cite{ref:c3bsw1}.

The integral $I_{-+}$ is obviously complex, the real part arising from the region where both $\pi$ and $K$ are off-shell and the imaginary part coming from the region where both $\pi$ and $K$ are on-shell. The details of the calculation are provided in the Appendix. The values of the mass scale $\Lambda$ are given in the next section. The final result is (to make the factorized amplitude real we drop a common factor of $i$ from the amplitudes in Eqs. (\ref{eq:c3decamp}) and (\ref{eq:c3interdec}))
\begin{eqnarray}
I _{-+}&=& 10^{-3} \tilde{G} \sqrt{{2\over 3}} F_0^{DK}(0) \left\{  8.735 + 1.769 {F_0^{D\pi}(0) \over F_0^{DK}(0)} +i(8.328 + 2.211 {F_0^{D\pi}(0) \over F_0^{DK}(0)}) \right\}  \nonumber \\
&=& |I_{-+}|exp(i\delta_I)~~GeV 
\label{eq:c3i}
\end{eqnarray}
Now, model predictions \cite{ref:c3fDPim} as well as experiments  \cite{ref:c3fDPie} give a ratio ${F_0^{D\pi}(0) \over F_0^{DK}(0)}\sim 1$. Consequently the phase $\delta_I \approx 45 ^{\circ}$ and it is insensitive to form factor-models. The magnitude of $I_{-+}$ depends on $F_0^{DK}(0)$. We use the value $F_0^{DK}(0)= 0.76$ \cite{ref:c3fDPim, ref:c3fDPie} to obtain
\begin{equation}
I_{-+} = 0.723 \times10^{-7} exp(i45^\circ)~~GeV.
\label{eq:c3i-+}
\end{equation}

\begin{figure}

%%Begin InstantTeX Picture
\let\picnaturalsize=N
\def\picsize{3in}
\def\picfilename{geta.eps}
%If you do not have the picture file add:
%\let\nopictures=Y
%to the beginning of the file.
\ifx\nopictures Y\else{\ifx\epsfloaded Y\else\input epsf \fi
\let\epsfloaded=Y
\centerline{\ifx\picnaturalsize N\epsfxsize \picsize\fi \epsfbox{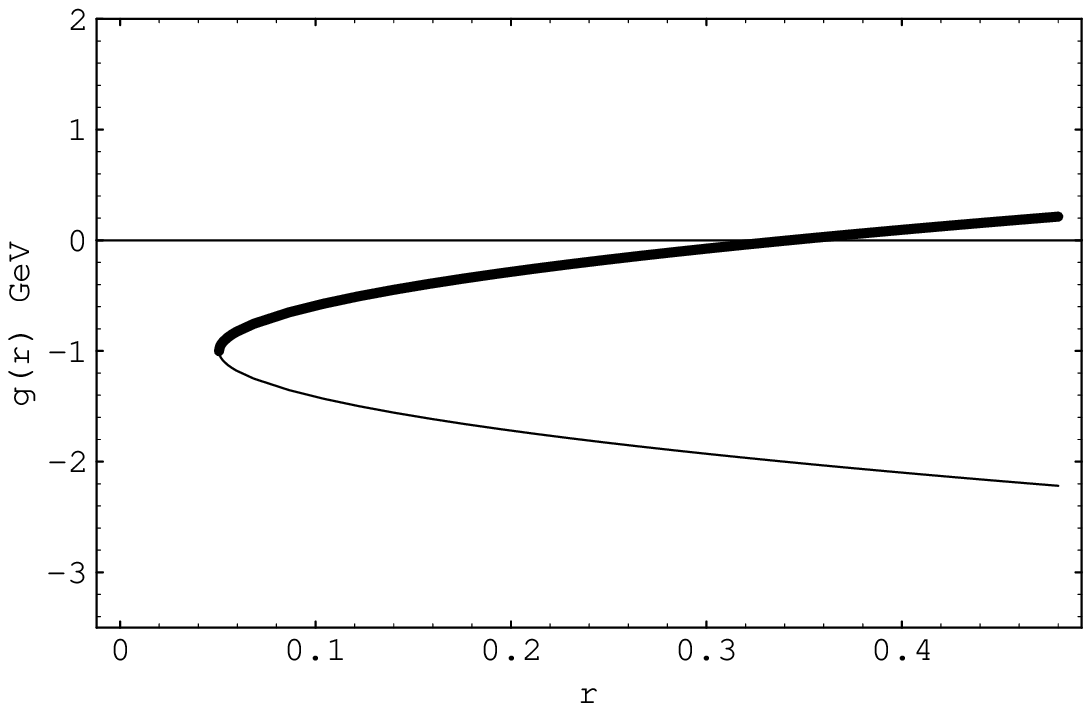}}}\fi
%%End InstantTeX Picture
\caption{Plot  for the two sets of solutions, $g^i_{\eta}(r), i=1,2 $, as a function of the branching ratio sum $r = Br(K^*_0(1950) \rightarrow K^0 \eta) +Br(K^*_0(1950) \rightarrow \bar{K}^0 \eta^\prime)$. The thick and light parts of the curve correspond to $g^1_{\eta}(r)$ and $g^2_{\eta}(r)$, respectively.}
\label{fig:c3geta}
\end{figure}

\begin{figure}
%%Begin InstantTeX Picture
\let\picnaturalsize=N
\def\picsize{3in}
\def\picfilename{getap.eps}
%If you do not have the picture file add:
%\let\nopictures=Y
%to the beginning of the file.
\ifx\nopictures Y\else{\ifx\epsfloaded Y\else\input epsf \fi
\let\epsfloaded=Y
\centerline{\ifx\picnaturalsize N\epsfxsize \picsize\fi \epsfbox{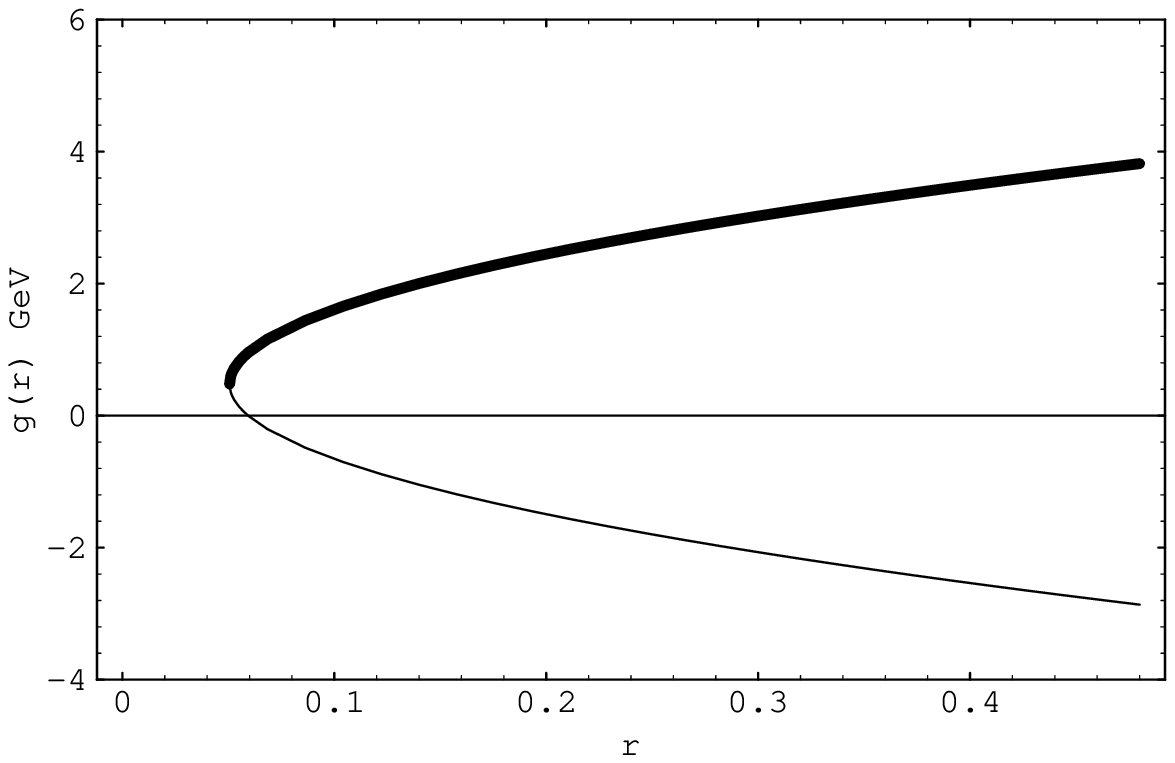}}}\fi
%%End InstantTeX Picture
\caption{Plot  for the two sets of solutions, $g^i_{\eta^\prime}(r), i=1,2 $, as a function of the branching ratio sum $r = Br(K^*_0(1950) \rightarrow K^0 \eta) +Br(K^*_0(1950) \rightarrow \bar{K}^0 \eta^\prime)$. The thick and light parts of the curve correspond to $g^1_{\eta^\prime}(r)$ and $g^2_{\eta^\prime}(r)$, respectively.}
\label{fig:c3getap}
\end{figure}

The resonant amplitude $A^r$ also depends on the strengths and the signs of strong coupling constants $f$ and $g$ which we determine as follows. The decay rate of a scalar particle decaying into two pseudoscalars is given by 
\begin{equation}
\Gamma(S \rightarrow P P) = { \mid {\bf p}\mid g^2_{SPP} \over 8\pi m_S^2}.
\label{eq:c3spp}
\end{equation}
For $\bar{K}^*_0(1950)$, we have the following measured branching ratio \cite{ref:c3pdg98}
\begin{equation}
s \equiv Br(K^*_0 \rightarrow K \pi) = 52 \pm 14 \%.
\label{eq:c3br}
\end{equation}
Using Eqs. (\ref{eq:c3g8}), (\ref{eq:c3spp}) and the central value of $s$ in Eq. (\ref{eq:c3br}), we get $g = \pm 2.707~ GeV$. 
Since the determination of $f \equiv g_{K_0^*\bar{K}^0\eta (\eta^\prime)}= g_{\eta(\eta^\prime)}$ is complicated by $\eta - \eta^\prime$ mixing and the fact that no measurements are available for the branching ratios $Br(K^*_0 \rightarrow K \eta)$ and $Br(K^*_0 \rightarrow K \eta^\prime)$, we provide some details of how we calculate $ g_\eta $ and $g_{\eta^\prime}$. We include $\eta - \eta^\prime$ mixing in the strong decay of the resonance as follows, 
\begin{eqnarray}
g_\eta &=& g_8\cos{\theta_p} -g_0\sin{\theta_p} \nonumber \\
g_{\eta^\prime} &=& g_8\sin{\theta_p} +g_0\cos{\theta_p},
\label{eq:c3gmix}
\end{eqnarray}
where the octet coupling $g_8$ is determined from Eqs. (\ref{eq:c3Hspp}) and (\ref{eq:c3g8}) to be
 $g_8 \equiv g_{K_0^*\bar{K}^0\eta_8 }= -g/\sqrt{6} $ and the unknown singlet coupling is $g_0 \equiv g_{K_0^*\bar{K}^0\eta_0}$.
We treat the following unmeasured branching ratio sum as a variable,
\begin{eqnarray}
r &=& Br(K^*_0(1950) \rightarrow K \eta) + Br(K^*_0(1950) \rightarrow K \eta^\prime) \nonumber \\
&=&{1 \over \Gamma_1 8\pi m_1^2}(|{\bf p}| g_\eta^2 + |{\bf p^\prime}| g_{\eta^\prime}^2),
\label{eq:c3brp}
\end{eqnarray}
where ${\bf p}$ and ${\bf p^\prime}$ are center of mass momenta of the final state particle in the decays $K^*_0 \rightarrow K \eta$ and $K^*_0 \rightarrow K \eta^\prime$, respectively. Since $g_8$ is known\footnote{The results of this paper were obtained using the positive value $g=2.707~GeV$. We learn nothing new if the negative value of $g$ is used. See Discussion.} from flavor-$SU(3)$ symmetry, we use Eqs. (\ref{eq:c3gmix}) and (\ref{eq:c3brp}) to solve for the singlet coupling $g_0(r)$ in terms of $r$. We obtain two solutions for $g_0$, denoted by $g_0^i(r), i=1,2$; we then substitute $g^i_0(r)$ in Eq. (\ref{eq:c3gmix}) and get two sets of solutions $(g^i_\eta(r),g^i_{\eta^\prime}(r)), i=1,2$. Their dependence on $r$ is shown in Figs. \ref{fig:c3geta} and \ref{fig:c3getap}.  In order for the strong coupling constants $g_\eta(r),g_{\eta^\prime}(r)$ to be real we find that we must have $r> 5 \%$. We also, have the constraint $r +s \leq 100 \%$, which restricts the allowed range for $r$ to:  $(5 \% \leq r\leq 52 \%)$. Using Eqs. (\ref{eq:c3a-+}) and (\ref{eq:c3g8}) with $g = 2.707 GeV$, the amplitude $A^{strong}_{-+}$ is found to be
\begin{equation}
A^{strong}_{-+}= - 5.45 g^i_{\eta (\eta^\prime)}(r) exp(i52^\circ).
\label{eq:c3astrong}
\end{equation}
Finally the total resonant amplitude $A^r$ is (in the following calculation all the amplitudes are scaled by a factor of $10^{-7}$),
\begin{eqnarray}
A_{-+}^r&=& I_{-+} \times A^{strong}_{-+} \nonumber \\ 
&=& -3.94g^i_{\eta (\eta^\prime)}(r) exp(i97^\circ)~~GeV,
\label{eq:c3ar-+}
\end{eqnarray}
and the total resonant amplitude $A^r$,
\begin{eqnarray}
A^r &=& {3 \over 2} A^r_{-+} \nonumber \\
& = & - 5.91 g^i_{\eta (\eta^\prime)}(r) exp(i97^\circ)~~GeV.
\label{eq:c3art}
\end{eqnarray}
Note that  the amplitude $A^{strong}$ defined in Eq. (\ref{eq:c3a-+}) is complex; however, the phase  of $A^r$ is not the phase of $A^{strong}$. The total amplitude for $D^0 \rightarrow \bar{K}^0\eta (\eta^\prime)$ is the coherent sum
\begin{equation}
A = A^f +  A^r.
\end{equation}

\bbc 
{\bf III. RESULTS}
\eec

For numerical calculations we used the following parameters, 
\begin{equation}
V_{cs} = 0.974,~~~~  V_{ud} = 0.975,~~~~ f_{\pi} = 0.131 ~GeV,~~~~ f_K = 0.160 ~GeV, 
\end{equation}
and the pole mass $\Lambda$,
\begin{equation}
m(sc(0^+))= 2.6 ~GeV, ~~~~ m(dc(0^+)) = 2.47 ~GeV. 
\label{eq:c3mlamb}
\end{equation}
The values of the form factors are presented in Table I.\\

\bbc
{\bf A. Factorized Amplitude} 
\eec

For the purpose of comparison with experiment we have calculated the factorized amplitude $A^f(D^0 \rightarrow K\eta (\eta^\prime))$ using the following models for the form factors: i) Bauer, Stech and Wirbel (BSW) model \cite{ref:c3bsw1}, where an infinite momentum frame is used to calculate the form factors at $q^2 = 0$, and a monopole form for $q^2$ dependence is 
assumed to extrapolate all the form factors to the desired value of $q^2$; ii) 
Casalbuoni, Deandrea, Di Bartolomeo, Feruglio, Gatto and Nardulli  
(CDDGFN) model \cite{ref:c3cdd}, where the form factors are evaluated at 
$q^2 = 0$ in an effective Lagrangian satisfying heavy quark spin-flavor 
symmetry in which light vector particles are introduced as gauge 
particles in a broken chiral symmetry.  A monopole form is used for the 
$q^2$ dependence. The experimental inputs for this model are from the 
semileptonic decay $ D \rightarrow  \pi l \nu$, and we have used the value 
of the form factors $F_1^{D\pi}(0) = F_0^{D\pi}(0) = 0.83 \pm 0.08$ \cite{ref:c3cckp} extracted from data, and decay constants $f_{D_s} = 213^{+14}_{-11} \pm 11,  f_{D} = 194^{+14}_{-10} \pm 10$ MeV\cite{ref:c3kadra} in calculating the weak coupling constants of the model at $q^2 = 0$ 
\cite{ref:c3cdd} , which are subsequently used in evaluating the required 
form factors ; vi) Isgur, Scora, Grinstein and Wise (ISGW) model 
\cite{ref:c3isgw}, where a non-relativistic quark model is used to 
calculate the form factors at zero recoil and an exponential $q^2$ 
dependence, based on a potential-model calculation of the meson  wave function, is use to extrapolate them to the desired $q^2$; iv) Lubicz, Martinelli, McCarthy and Sachrajda (LMMS) model \cite{ref:c3LMMS}, where the form factors are obtained from lattice calculation of  $D$ meson semileptonic decays.
\\

The factorized amplitudes with model form factors and the experimentally determined amplitudes are presented in Table II. The prediction for $D^0 \rightarrow K\eta$ amplitude is too low in every case; an enhancement of a factor of $1.5$ to $3$, depending on the model, is needed to match the experimental amplitude. For $D^0 \rightarrow K\eta^\prime$, the situation is worse. \\

\bbc
{\bf B. Resonant Amplitude} 
\eec

In the following we list the amplitude represented by the diagram of Fig. \ref{fig:c3.2} separately for the cases where the loop particles are on-shell and off-shell.\\

\bbc
{\it 1. On-shell contribution:}
\eec

The contribution to the resonant amplitude $A^r$ from on-shell loop particles is obtained by taking the imaginary part of the integral $I_{-+}$. We get from eq. (\ref{eq:c3arsym}), (\ref{eq:c3i-+}) and (\ref{eq:c3astrong}),
\begin{eqnarray}
A^r (on\mbox{-}shell) &=& { 3\over 2} i Im(I_{-+}) \times A^{strong}_{-+} \nonumber \\
&=&-4.18 g^i_{\eta(\eta^\prime)}(r) exp(i142^\circ) ~GeV.
\label{eq:c3ashel}
\end{eqnarray} 

\bbc
{\it 2. Off\mbox{-}shell contribution:}
\eec

\begin{figure}

%%Begin InstantTeX Picture
\let\picnaturalsize=N
\def\picsize{3.5in}
\def\picfilename{atotaleta3.eps}
%If you do not have the picture file add:
%\let\nopictures=Y
%to the beginning of the file.
\ifx\nopictures Y\else{\ifx\epsfloaded Y\else\input epsf \fi
\let\epsfloaded=Y
\centerline{\ifx\picnaturalsize N\epsfxsize \picsize\fi \epsfbox{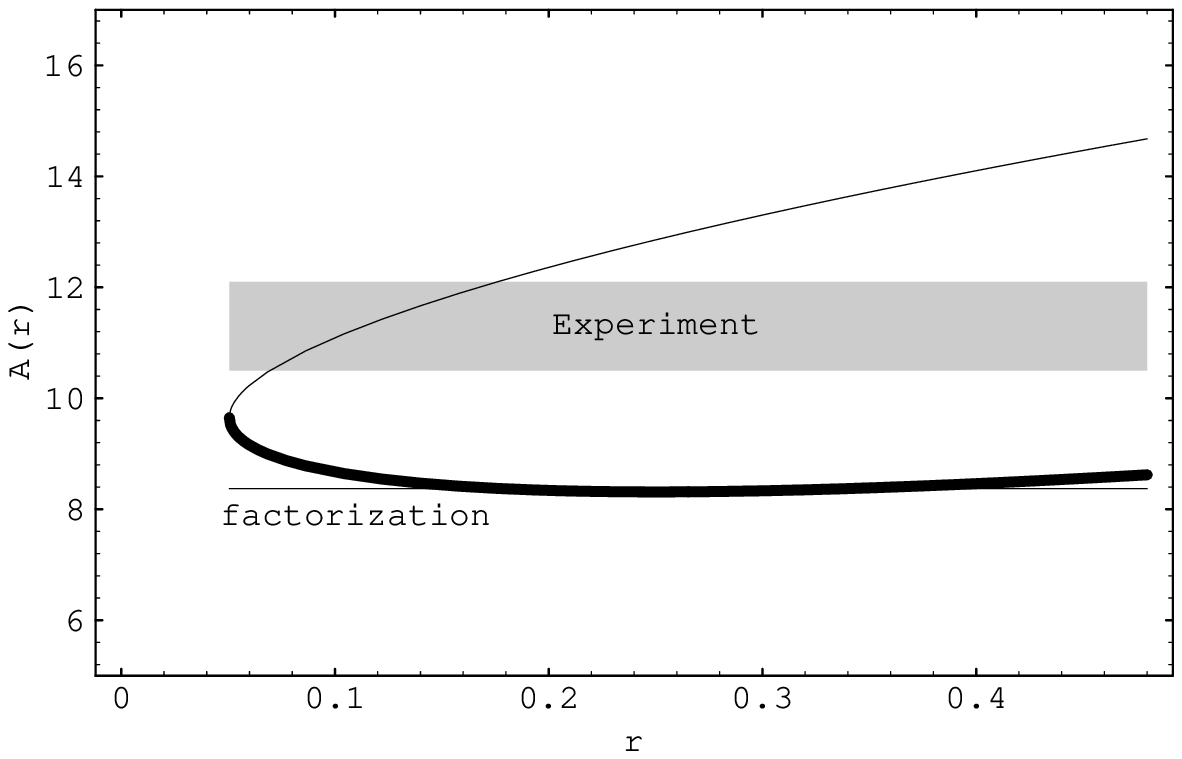}}}\fi
%%End InstantTeX Picture
\caption{Plot of the magnitude of the total amplitude $|A| = |A^f + A^r|$ in Eq. (\ref{eq:c3at}) for $D^0 \rightarrow \bar{K}^0 \eta$  as a function of the branching ratio sum $r = Br(K^*_0(1950) \rightarrow \bar {K}^0 \eta) +Br(K^*_0(1950) \rightarrow \bar{K}^0 \eta^\prime)$. The thick and light parts of the curve correspond to the solution $g_{\eta}^1$ and $g_{\eta}^2$, respectively. The shaded region represents the experimental value of the amplitude, the horizontal line represents the factorized amplitude $A^f$. The values on the y-axis must be multiplied by a factor of $10^{-7} ~GeV$ to get the absolute magnitude of the decay amplitude. }
\label{fig:c3ateta}
\end{figure}

\begin{figure}

%%Begin InstantTeX Picture
\let\picnaturalsize=N
\def\picsize{3.50in}
\def\picfilename{atotaletap3.eps}
%If you do not have the picture file add:
%\let\nopictures=Y
%to the beginning of the file.
\ifx\nopictures Y\else{\ifx\epsfloaded Y\else\input epsf \fi
\let\epsfloaded=Y
\centerline{\ifx\picnaturalsize N\epsfxsize \picsize\fi \epsfbox{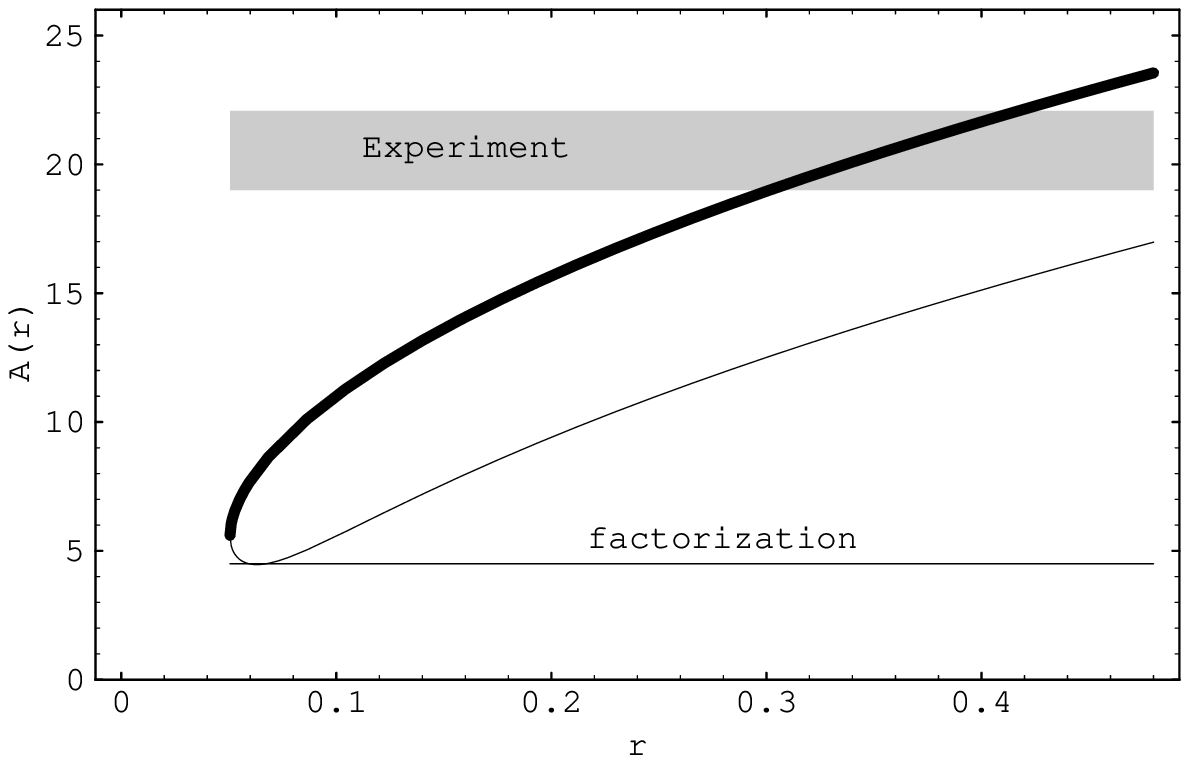}}}\fi
%%End InstantTeX Picture
\caption{Plot of the magnitude of the total amplitude $|A| = |A^f + A^r|$ in Eq. (\ref{eq:c3at}) for $D^0 \rightarrow \bar{K}^0 \eta^\prime$  as a function of the branching ratio sum $r = Br(K^*_0(1950) \rightarrow K^0 \eta) +Br(K^*_0(1950) \rightarrow \bar{K}^0 \eta^\prime)$. The thick and light parts of the curve correspond to the solution $g_{\eta^\prime}^1$ and $g_{\eta^\prime}^2$, respectively. The shaded region represents the experimental value of the amplitude, the horizontal line represents the factorized amplitude $A^f$. The values on the y-axis must be multiplied by a factor of $10^{-7} ~GeV$ to get the absolute magnitude of the decay amplitude. }
\label{fig:c3atetap}
\end{figure}

For the loop particles off-shell the resonant amplitude $A^r$ is obtained by taking the real part of the integral $I_{-+}$. We get,
\begin{eqnarray}
A^r(off \mbox{-}shell) &=& { 3\over 2}Real(I_{-+}) \times A^{strong}_{-+} \nonumber \\
&=&-4.18 g^i_{\eta(\eta^\prime)}(r) exp(i52^\circ) ~ GeV.
\label{eq:c3aofshel}
\end{eqnarray}
Note that the on-shell contribution, Eq. (\ref{eq:c3ashel}), is of the same size as the off-shell, Eq. (\ref{eq:c3aofshel}), but advanced by a phase of $90^\circ$ which comes from the factor $i$. 
The on-shell and off-shell, amplitudes have the same magnitude but different phases, therefore including off-shell effect  will modify both the amplitude and the phase of the resonant amplitude.
The total resonant amplitude $A^r$ is given by,
\begin{equation}
A^r = A^r(on\mbox{-}shell) + A^r(off\mbox{-}shell).
\label{eq:c3onoff}
\end{equation}
Finally, the total decay amplitude is obtained by adding the factorized amplitude $A^f$ to the resonant amplitude $A^r$,
\begin{equation}
A = A^f  - 5.91 g^i_{\eta(\eta^\prime)}(r) exp(i97^\circ) ~ GeV.
\label{eq:c3at}
\end{equation}
Plots of the magnitude $|A| = |A^f + A^r|$ as a function of $r$ are given in Figs. \ref{fig:c3ateta} and \ref{fig:c3atetap} for the decays $D \rightarrow K \eta $ and $D \rightarrow K \eta^\prime $, respectively. In these figures we have used $A^f(D^0 \rightarrow K \eta) = 8.37~GeV$ and $A^f(D^0 \rightarrow K \eta) = 4.5~GeV$ as predicted in $BSW$ model only. We discuss the results in the next section. \\

\bbc
{\it 3. Strong phases}
\eec

The phase of the total resonant amplitude and the on-shell amplitude can be determined from Eqs. (\ref{eq:c3art}) and (\ref{eq:c3ashel}), respectively. The sign of the coupling constant, $g_{\eta}^i$ and $g_{\eta\prime}^i$ is important; a change in sign introduces a phase difference of $180^\circ$ which leads to a different pattern of interference (constructive or destructive) between the factorized amplitude $A^f$ and the resonant amplitude $A^r$.  We use the graphs of Figs. \ref{fig:c3geta} and \ref{fig:c3getap} to read the signs of $g_{\eta}(r)$ and $g_{\eta^\prime}(r)$ and then determine the strong phase of the resonant amplitudes. The phases are summarized in Table III. \\

\bbc
{\bf IV.  DISCUSSION}
\eec

The factorization prediction for the amplitude $A(D^0 \rightarrow \bar{K}^0\eta)$ is too low compared to the experimental data and the situation is even worse for $A(D^0 \rightarrow \bar{K}^0\eta^\prime)$ (see Table II).\\

Amplitudes with the iclusion of resonant $FSI$ (Eq. \ref{eq:c3at}) are plotted in Figs. \ref{fig:c3ateta} and \ref{fig:c3atetap} for $A(D^0 \rightarrow \bar{K}^0\eta)$ and $A(D^0 \rightarrow \bar{K}^0\eta^\prime)$, respectively. From Fig. \ref{fig:c3ateta} we notice that for solution $i = 1$, $A(D^0 \rightarrow \bar{K}^0\eta)\sim A^f(D^0 \rightarrow \bar{K}^0\eta)$ over most of the range of $r$, except where $r$ is low, $r\leq 14 \%$, when 
$A^f(D^0 \rightarrow \bar{K}^0\eta)$ gets a small enhancement over its factorized value, but stays below the experimental value. On the other hand, Fig. \ref{fig:c3atetap} shows that for solution $i = 1$, $A(D^0 \rightarrow \bar{K}^0\eta^\prime)$ rises with $r$ and fits the experiment in the range $30\% \leq r \leq 42 \%$. Although, for lower values of $r$ $A(D^0 \rightarrow \bar{K}^0\eta^\prime)$ is underestimated, it still gets large enhancement over its factorized value. Solution $i = 1$ does not accomodate a simultaneous fit to $A(D^0 \rightarrow \bar{K}^0\eta)$ and $A(D^0 \rightarrow \bar{K}^0\eta^\prime)$.

As for solution $i = 2$, we notice from Figs. \ref{fig:c3ateta} and \ref{fig:c3atetap} that both $A(D^0 \rightarrow \bar{K}^0\eta)$ and $A(D^0 \rightarrow \bar{K}^0\eta^\prime)$ rise with $r$. In particular, a fit to $A(D^0 \rightarrow \bar{K}^0\eta)$ is secured for $6\% \leq r \leq 20 \%$. Despite the fact that in this range of $r$ $A(D^0 \rightarrow \bar{K}^0\eta^\prime)$ could be raised by almost $100\%$ over its factorized value, it still remains well below its measured value. Again a simultaneous fit to $A(D^0 \rightarrow \bar{K}^0\eta)$ and $A(D^0 \rightarrow \bar{K}^0\eta^\prime)$ is eluded.    \\         
      
Recently Dai et al. \cite{ref:c3dai} have used the same mechanism, but kept only the on-shell loop contribution to estimate the effect of resonant $FSI$ in the decay $D^0 \rightarrow K^0 \bar{K}^0$. Hence, contrary to our case, the strong phase is solely determined by the resonant propagator.\\

In a recent paper Gronau \cite{ref:c3gro} has calculated the contribution of $K^*_0(1430)$ in the direct channel (annihilation topology, as in our case) to $D^0 \rightarrow K^- \pi^+$ decay in a model-independent way and found it to be a substantial fraction $(\sim 20 \%)$ of the total amplitude (which is largely isospin 1/2). He argues that the effect of $K^*_0(1950)$ on $D^0 \rightarrow K^- \pi^+$ decay could even be larger. The difference between his approach and ours (apart from the fact that we are dealing with different $D^0$ decay modes) is that we have used the factorized input for the weak transition $D^0 \rightarrow \bar{K}^*_0(1950)$ through a $\pi K$ loop, while Ref. \cite{ref:c3gro} uses current algebra, with smoothness assumption, to relate $D^0 \rightarrow \bar{K}^*_0(1430)$ vertex to $D^+ \rightarrow \bar{K}^*_0(1430) \pi^+$ measured rate. Thus while the $D^0 \rightarrow \bar{K}^*_0(1950)$ vertex of Ref. \cite{ref:c3gro} is assumed to be real, ours is complex. If our resonant contribution is equated to the W-exchange amplitude of Ref. \cite{ref:c3gro}, then clearly the resonant contribution has a phase around $90^\circ$ (modulo $\pi$) relative to the tree amplitude (see Eq. (\ref{eq:c3at})).

In summary we find that the resonant $FSI$ due to $K^*_0(1950)$ in the direct channel effects $A(D \rightarrow \bar{K}^0 \eta)$ and $A(D \rightarrow \bar{K}^0 \eta^\prime)$ significantly. However, a simultaneous fit to both decay amplitudes is not possible.   

A final comment: We tried the same calculation with a negative sign for $g$, $g = -2.707~GeV$. Apart from leading to a phase shift of $180^\circ $ in the strong phase $\delta_r$, it did not change our conclusions. It was still impossible to fit both $A(D \rightarrow \bar{K}^0 \eta)$ and $A(D \rightarrow \bar{K}^0 \eta^\prime)$ simultaneously.  

\bbc 
{\bf Appendix}
\eec
Because of Eq. (\ref{eq:c3arsym}), the calculation of the resonant amplitude $A^r$ reduces to that of $A_{-+}^r = I_{-+}A^{strong}_{-+}$. Using the vertex function given in Eq. (\ref{eq:c3vertex}) in Eqs. (\ref{eq:c3ar-+}) and (\ref{eq:c3a00})  we get the following integral \\
\begin{eqnarray}
I_{-+} &=&{-i \tilde{G} \over (2\pi)^4} \sqrt{{2 \over 3}} \int {d^4k \over (k^2 - m_{\pi}^2)((w-k)^2 - m_K^2)}  \times \nonumber \\
 & &\left\{\sqrt{{2 \over 3}}a_1 f_{\pi}(m_D^2 - (w-k)^2)F_0^{DK}(k^2) - {a_2 \over \sqrt{6} }  f_K(m_D^2 - k^2)F_0^{D\pi}((w - k)^2)\right\},
\label{eq:c3ba}
\end{eqnarray}
where $d^4k = dk_0d^3{\bf k}$, $ k^2 = k_0^2 - |{\bf k}|^2$, and $w =(m_D,0)$. Using form factors as explained in the text, the integral $I_{-+}$ reduces to the following form,
\begin{eqnarray}
I_{-+} &=&{- \tilde{G}\over (2\pi)^4}\sqrt{{2 \over 3}}  \left\{  a_1 \sqrt{{2 \over 3}}\Lambda^4 f_{\pi}F_0^{DK}(0)\left[ (m_D^2 - m_K^2)I_1^{DK} - I_2^{DK} \right] \right. \nonumber \\
 & &\left. - a_2 {f_K \over \sqrt{6}}\Lambda^4F_0^{D\pi}(0)\left[ (m_D^2 - m_{\pi}^2)I_1^{D\pi} - I_3^{D\pi} \right] \right\},
\end{eqnarray}
where the integrals $I_i$ are given by
\begin{eqnarray}
I_1 &=&{1 \over (\Lambda^2 - m_\pi^2)(\Lambda^2 - m_K^2)} \left( I_{\Lambda\Lambda} -I_{\Lambda K} -I_{\pi\Lambda} + I_{\pi K}  \right), \\
I_2 &=&{-1 \over (\Lambda^2 - m_\pi^2)} I_{\pi\Lambda} + {1 \over (\Lambda^2 - m_\pi^2)}I_{\Lambda\Lambda}, \\
I_3 &=&{-1 \over (\Lambda^2 - m_k^2)} I_{\Lambda K} + {1 \over (\Lambda^2 - m_k^2)}I_{\Lambda\Lambda}.
\end{eqnarray}
The integrals $I_i \equiv I_i^{DK(\pi)}$ with $\Lambda$ having the appropriate mass given in Eq. (\ref{eq:c3mlamb}).
The integral $I_{XY}$ has the generic form
\begin{equation}
I_{XY} = i \int{{dk_0d^3{\bf k} \over (k^2 - m_X^2)((w-k)^2 - m_Y^2)}}.
\end{equation}
First we integrate over $dk_0$ in the complex plane with a contour closed in the lower half plane. We have used a mass scale Eq. (\ref{eq:c3mlamb}) $\Lambda > m_{D^0} + m_K$, therefore the integration over $d^3{\bf k}$ is well defined for all the integrals except $I_{\pi K}$ which has a pole contributing to the imaginary part of $I_{-+}$. The above calculation leads to the following result, 
\begin{eqnarray}
I_{-+} &=& \sqrt{{2 \over 3}}\left\{8.735 F_0^{DK}(0) + 1.769 F_0^{D\pi}(0) \right. \nonumber \\ 
&&\left.+i(8.328 F_0^{DK}(0) + 2.211 F_0^{D\pi}(0))\right\}\times10^{-3} \tilde{G} ~~GeV.
\end{eqnarray}
\\

%{{{{{{{{{{{{{{{{{{{{{{{{{{{{{{{{{{{{{{}}}}}}}}}}}}}}}}}}}}}}}}}}}}}}}}}}}}}

%+++++++++++++++++++++++++++++++++++++++++++

%(((((((((((((((((((((((((((()))))))))))))))))))))))))))))))))))))

\begin{table}
\centering
\caption{Model  predictions for the form factors : $F_0^{D\eta (\eta^\prime)}(m_K^2)$,
$F_0^{DK (\pi)}(0)$ and the ratio ${ F_0^{D\pi}(0)\over F_0^{DK}(0)}$ for the processes $D^0 \longrightarrow \bar{K}^0 \eta, \bar{K}^0 \eta^\prime, K\pi  $.}
\vspace{5 mm}
\begin{tabular}{|c|c c c c|} \hline
&$BSWI$&$CDDFGN$&$ISGW$&$LMMS$ \\ \hline
$F_0^{D\eta}(m_K^2)$&0.710&0.313&0.638&0.344 \\
$F_0^{D\eta ^{\prime}}(m_K^2)$&0.683 &-&0.937&0.240 \\ 
$F_0^{DK}(0)$&0.762&0.699&0.769&0.63 \\
$F_0^{D\pi}(0)$&0.692 &0.83&0.510&0.58 \\ 
${ F_0^{D\pi}(0)\over F_0^{DK}(0)}$& 0.91&1.19&0.66&0.92\\  \hline
\end{tabular}
\label{tab:c3resff}
\end{table} 

%---------------------------------------------

\begin{table}
\centering
\caption{Model  predictions for the factorized amplitude $A^f$ for the process $D^0 \longrightarrow \bar{K}^0 \eta (\eta^\prime)$ These values must be multiplied by a factor of  $10^{-7}GeV$.}
\vspace{5 mm}
\begin{tabular}{|c|c c c c c|} \hline
&$BSWI$&$CDDFGN$&$ISGW$&$LMM$&$Expt.\cite{ref:c3pdg98}$ \\ \hline
$A^f(D^0 \longrightarrow \bar{K}^0 \eta )$&8.37&3.70&7.53&4.06&$11.3\pm 0.8$ \\
$A^f(D^0 \longrightarrow \bar{K}^0 \eta^\prime )$&4.50 &-&6.18&1.58&$20.54\pm 1.54$ \\ \hline
\end{tabular}
\label{tab:c3resrate}
\end{table} 

%------------------------------------------

\begin{table}
\centering
\caption{Phases of the total resonant and on-shell amplitudes defined in Eqs. (\ref{eq:c3art}) and (\ref{eq:c3ashel}), respectively, as functions of $r$ for the processes $D^0 \longrightarrow \bar{K}^0 \eta (\eta^\prime)$. $\delta_r$ is the phase of $A^r$ (Eq. (\ref{eq:c3art})) and $\delta_{on\mbox{-}shell}$ is the phase of $A^r(on\mbox{-}shell)$ (Eq. (\ref{eq:c3ashel})).}
\vspace{5 mm}
\begin{tabular}{|c|c|c c c|} \hline
Solution&Decay&$r$&$\delta_r$&$\delta_{on\mbox{-}shell}$ \\ \hline
&$D^0 \longrightarrow \bar{K}^0 \eta $&$(5 \leq r \leq 35)\%$&$97^\circ $&$142^\circ $ \\ 
$i = 1$&&$(35 \leq r \leq 48)\%$&$(97 \pm 180)^\circ$&$(142 \pm 180)^\circ $ \\ 
&$D^0 \longrightarrow \bar{K}^0 \eta^\prime $&$(5 \leq r \leq 48)\%$&$(97 \pm 180)^\circ$&$(142 \pm 180)^\circ$ \\ \hline
$i = 2$&$D^0 \longrightarrow \bar{K}^0 \eta $&$(5 \leq r \leq 48)\%$&$97^\circ $&$142^\circ $ \\ 
&$D^0 \longrightarrow \bar{K}^0 \eta^\prime $&$(5 < r \leq 48)\%$&$97^\circ$&$142^\circ$ \\ \hline
\end{tabular}
\label{tab:c3phase1}
\end{table}

\vspace{0.25cm}
{\bf Acknowledgments}:
We appreciate computational  help provided by Dr. F. M. Al-Shamali. This research was partially funded by the Natural Sciences and Engineering Research Council of Canada through a grant to A. N. K.

\pagebreak

\end{document}